# Applicability of Overlay Non-Delay Tolerant Position-based Protocols in Highways and Urban Environments for VANET

Mahmoud Ali Al Shugran[1]

***ABSTRACT***

*Vehicular Ad hoc Network (VANET) is a new sort of wireless ad-hoc network. Vehicle-to-Vehicle (V2V) communication is one of the main communication paradigms that provide a level of safety and convenience to drivers and passengers on the road. In such an environment, routing data packets is challenging due to frequent changes of network topology because of the highly dynamic nature of vehicles. Thus, routing in VANETs requires efficient protocols that guarantee message transmission among vehicles. Numerous routing protocols and algorithms have been proposed or enhanced to solve the aforementioned problems. Many position-based routing protocols have been developed for routing messages that have been identified to be appropriate for VANETs. This work explores the performances of selected unicast non-delay tolerant overlay position-based routing protocols. The evaluation has been conducted in highway and urban environments in two different scenarios. The evaluation metrics that are used are Packet Delivery Ratio (PDR), Void Problem Occurrence (VPO), and Average Hop Count (AHC).*

## 1. Introduction

Vehicular ad hoc network or Vehicular network (VANET) is a distributed self–configure, self-routing and self–organize network with no need for administrator. Vehicles are the cornerstone of VANET. Thus, VANET networks can set up routing paths along vehicles. The VANET is implemented on city, urban and highway environments. In such environments, VANET networks serve many safety and non-safety applications. The most challenging matter in VANETs is the routing of data packets. This is because the frequent varying of vehicles' speed that incurs frequently topology changing. That results in breakage of link communication between vehicles. Also, the variable of vehicles' velocity contributes in a sparse network some time and some time dense network. Hence, such type of scenarios changes the performance of the VANET network very frequently [1, 2, 3].

In VANET, there are three types of communications are involved viz. Communication between the vehicles (V2V), Vehicle to infrastructure communication (V2I) and cellular communication. The V2V routing protocols on the basis of routing information's and transmission strategies are broadly categorized into unicast, multicast, broadcast [4]. The concern of this work is about unicast category that is further subdivided into topology based and position based. Traditional routing protocols related to this category are unable to fully address the new features and requirements of VANETs. Firstly, the existing topology based or position based protocols are designed for efficient routing with low latency and no congestion. Secondly these protocols are designed for specific scenarios. Since dynamic topology of VANET conflict the routing algorithm and at run time the network may attain any kind of scenario. To solve the routing challenges in the vehicular networks, a numerous protocols are developed for VANETs. Researchers found that position-based routing protocols are most convenient for VANET than others. Thus, the concern of this paper focuses on investigating the available unicast non-delay tolerant overlay position-based routing protocols.

The rest of the paper is organized as follows: Section 2 devoted to VANET communication environment and applications. Section 3 discusses the routing issues in VANETS. Section 4 presents unicast routing protocols in VANET. Section 5 present detailed representative Non-DTN Overlay position-based routing protocols. Next, section 6 shows the performance evaluation of the selected protocols by giving a particular interest to specific performance metrics. Section 7 presents experimental results and discussion and finally, section 8 conclude this study and suggest directions for future work.

## 2. VANET Communication Environment and Applications

VANET is not restricted to space like indoor or outdoor. VANETs implemented in two different environments; Highways environment and Rural environment. Communication on each environment can be either vehicle to vehicle (V2V) communication or vehicle to infrastructure on road-unit (V2I communication. Thus, the intended routing algorithm that will be used in VANET must adapt to the environment that will be deployed in and the communication type [5]. In general VANET applications can be characterized on the basis of their functionalities; they can be broadly categorized as safety and non-safety applications for the sake of commercial and convenience. Safety applications primarily focus on passenger safety to mitigate the opportunity of road accidents by avoiding intersection collision. Furthermore, safety applications alert for the curve speed, cooperative forward collision alert, traffic signal and other applications. The non-safety applications enlarge the on road passengers' satisfaction by providing different facilities such as social media. Further, non-safety applications provide electronic toll collection system that allows driver to pay toll online and preventing time wasting. Also, the non-safety applications can provide parking lot payment, traffic management and many others services. Compared to safety applications, the non-safety applications consume high bandwidth and require significant network resources [6].

There are several important differences between an urban and a highway environment. These differences have a major impact on communication in each environment that imposes critical demand in designing an efficient routing protocol. In an urban environment, there are many obstacles such as vehicles themselves, buildings, corners, and junctions that affect signal propagation. Hence, the routing protocol for the urban environment should provide a mechanism to avoid obstacles. On the other hand, signals propagate smoothly in the absence of obstacles in the highway environment. Furthermore, from a routing point of view used routing algorithm can perform several options to forward packets in an urban environment because there are many streets, avenues that are close to each other. Thus, a driver can drive straight-ahead or the driver can turn to a different road. Hence, the routing protocol for the urban environment should maintain the nodes' neighbourhood always update. On the other hand, on a highway, there are only a few entrances and exits and no crossroads; therefore, most of the time the vehicles can only go forward. Thus, the routing algorithm has a very limited option in forwarding the packets. Besides, the speed of vehicles and traffic lights has an important impact on node density and connectivity. Since each environment has different condition and because speed inside urban is low (usually limited to 50km/h) the connectivity and node density is high. On the other hand, in a highway the speed is almost high about 120km/h, hence, the connectivity and node density is low. Based on the differences between both environments, my study investigates if one routing protocols can be efficiently used in both environments [7, 8].

## 3. Routing Issues in VANET

In VANET, invention an efficient routing protocol that is satisfy VANET requirements is one of the essential challenge that need to be investigated. Designing an efficient protocol for VANETs is very crucial. As aforementioned, VANET challenges and its new characteristics compared to other wireless networks are the main area of work for routing efficiency. Furthermore, the Intelligent Transportation system (ITS) faces many challenges in routing paradigm for VANET. The cornerstone of developing efficient routing protocol in VANET is to achieve successful communication between nodes with a short rate of dropped packets and provide a minimal amount of the overhead. The intended routing protocols for the VANETs must take into account the highly dynamic topology. It's obvious that, applying existing routing protocols in vehicular networks is ineffectual. Therefore, modifying or enhancing the existing protocols is the usual requirement to resolve the routing challenge in the VANETs or establishing new routing protocols that satisfy the new VANET requirements. The specific features of VANET that they are very dynamic environment since they are formed with vehicles that join and leave the network very fast [5, 6].

As aforementioned, VANET possess some particular characteristics making them different from the other wireless ad hoc networks in several aspects. In a brief the VANETs are characterized firstly by their quick changes in network topology. Secondly, the link lifetime between vehicles is very short mainly because of the Vehicles high speed in highway area and low density of vehicles in urban area. Thirdly, there are different environment where VANET can be deployed that imposes several requirements. Finally, there are different applications that are run over the VANET which make it a unique environment that required special demand to be satisfied [5, 6].

## 4. Unicast Routing Protocols in VANET

In VANET, efficient routing protocols have to deploy unique designs that guarantee reliable communication between two communicating nodes with no disruption. Vehicular ad hoc networks reinforce several communication patterns that can be classified into three categories Unicast communication, Multicast/ Geocast communication and Broadcast communication. The concern of this work focuses on Unicast approach. Unicast communication aims to send a data packet from a single node to another single node. The targeted node can be reached via multi-hop wireless communication. Some routing protocols that use the unicast paradigm may also need to use a multicast paradigm too [6, 7, 8]. The main aim of a wireless ad hoc network routing algorithm is to guarantee reliable communication between two communicating nodes correctly and efficiently according to the expected QoS parameters. The formation of a route should satisfy the bandwidth consumption with minimum overhead. This section will discuss the two main categories of VANET routing protocols [6, 7, 8].

As shown in Figure 1 there are two well-known categories for data packet forwarding commonly adopted in multi-hop wireless ad hoc networks for VANETs which are topology-based and position-based routing protocols [9, 10, 11, 12]. Topology-based routing uses global information about the network topology and the information about the communication links for making routing decisions. In this case, every node maintains a routing table, which is the case of routing protocols for MANETs. Position-based routing uses neighbouring location information to perform packet forwarding [9, 10, 11, 12].

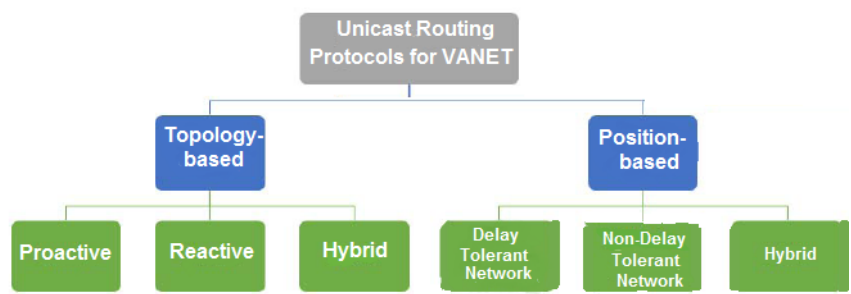

Figure 1. Data packet forwarding approaches for VANETs

### 4.1. Topology Based Routing Protocols

Topology based protocols utilize the information about links that exist in the network to perform data packet forwarding. In this case, every node maintains a routing table. These protocols use either proactive (table-driven), reactive (on-demand) and hybrid approaches for routing. In proactive approaches each vehicle has a routing table that should be updated frequently to have a route for each other node in the network at all times. In VANET, the network topology changes frequently, and because the nodes should be up to date with such changes, this may occupy a significant part of the bandwidth. With reactive routing protocols routes are established only on demand. Hence, an additional delay is required at the beginning of each demand to find the rout. The hybrid protocols combine both reactive and proactive protocols. Hybrid protocols aim to minimize routing delays and overhead in the route discovery for which the nodes are divided into zones. These protocols are not effective in scenarios with high dynamic vehicles. The study by several researchers expose that topology based routing protocols are impractical for VANETs either due to the aforementioned reasons. All topology based routing protocols are important but the concern of this paper focuses on available unicast position-based routing protocols [9]-[12].

### 4.2. Position Based Routing Protocols

Several unicast routing protocols relies on position-based greedy forwarding strategy to provide vehicle -to- vehicle communication. In position-based routing approach the movement of nodes in VANET is limited to the road therefore; routing process using location information makes sense. The major prerequisite for position-based routing is that a sender node can obtain the current position of itself, neighbouring nodes as well as the destination through location service. Thus, the routing decisions require location services such as the Global Position System (GPS) device to determine the location of the participating nodes. Through using GPS, every vehicle can continuously send beacon packets with its position information. Position based protocols are suitable for VANETs since they offer a higher delivery ratio than topology-based routing protocols in a highly mobile environment. They provide a minimum delay in establishing the route and achieve good scalability compared to topology-based routing protocols. Three procedures recognize the operation of position-based routing protocols which are; Path selection, forwarding mechanism, and recovery strategy. Using a path selection algorithm is not mandatory in position-based routing. This algorithm can be used with the routing protocol if it presents more benefits to it. On the other hand, if the path selection algorithm is not used, there will be no extra overhead but the misleading of packets may occur [9]-[12].

There are several packet-forwarding strategies used for position-based protocols viz. trajectory-based forwarding, contention-based forwarding, opportunistic forwarding, greedy forwarding, and hybrid forwarding. The concern of this paper is the greedy forwarding mechanism. There are three main greedy-forwarding strategies: the traditional greedy approach, improved greedy, and restricted greedy. Most current position-based protocols use a greedy forwarding mechanism to route packets from a source to the destination. When applying a greedy algorithm, a source vehicle includes the position of the data packet' destination and selects the next-hop depending on the optimized criteria of the algorithm; for example the closest neighbour to the destination. Similarly, each intermediate vehicle selects a next-hop vehicle until the packet reaches the destination. To accomplish this, the vehicles should be able to periodically broadcast small packets (called beacons) with random jitter to announce their position and enable other vehicles to maintain a one-hop neighbour table. With high mobility in VANET, the beaconing frequency can be adapted to the degree of mobility. Even though, a fundamental problem of inaccurate position information is always present in VANET. A neighbour vehicle that is selected as a next-hop may no longer be in transmission range. This leads to a significant decrease in the packet delivery rate. To improve the accuracy of position information, it is possible to increase the beaconing frequency. However, Periodic beaconing creates a lot of congestion collision probability. The recovery strategy defines the actions that a node must perform when it does not have any neighbour that meets the forwarding criteria. Based on the recovery methods position-based routing protocols are subdivided into Delay Tolerant Network (DTN), Non-Delay Tolerant

Network (Non-DTN), and Hybrid [9]-[12].

### 4.2.1. The DTN position-based Routing protocols

Routing with this strategy deploys the carry and forwarding recovery mechanism where the sender node holds a packet in the cache until it finds a suitable next forwarding node. The carry and forward mechanism aim to transfer the data packets between the nodes even with the absence of the path between the communicating vehicles. The carry and forwarding recovery mechanism needs extra architecture. The DTN strategy introduces a high delay in data transmission. But this strategy is suitable for delay-tolerant applications. Thus, the DTN recovery approach has increased its use in VANET routing protocols. As another position-based routing protocol Vehicular Delay Tolerant Network (DTN) uses geographical knowledge needed by the routing protocols [9]-[12]. The Vehicle-Assisted Data Delivery (VADD) [13] and (GeOpps) [14] Geographical Opportunistic Routing are well-known routing protocol belong to DTN.

### 4.2.2. The Non-DTN Position-Based Routing protocols

Routing with this strategy has the ability to change the forwarding criteria to find a suitable node to forward the packet to avoid carrying it through using greedy forwarding approaches. Therefore, Non-DTN protocols aim to decrease the delay at packets delivery from the source node to the destination node. The Non-DTN protocols are appropriate to be used with critical safety applications, which demand real-time response during data dissemination. Hence, the delay time in the transmission is the main concern when designing a Non-DTN protocol. To satisfy this demand, the shortest path method is usually adopted. Nevertheless, the shortest path sometimes introduces some delay because of holes in the network. Many types of NON-DTN protocols have been developed to handle this failure, namely, beacon, no-beacon, and hybrid protocols. The concern of this work concentrates on the beacon approach which subdivided into two categories non-overlay and overlay network [8]-[12].

**4.2.2.1. Non-Overlay**: The Non-overlay forwarding mechanism takes routing decisions at each hop. All protocols in the Non-Overlay network use the greedy forwarding algorithm for sending data from source vehicle to destination vehicle. When no neighbour is closer to the destination other than the current one the greedy forwarding method can fail. In this case, the recovery strategy will be applied. One of the earlier routing protocols that apply the non-overlay approach is the Greedy Perimeter Stateless Routing (GPSR) [15].

**4.2.2.2 Overlay**: An overlay network, logical links that connect all nodes are built on the top of another overlay network to get information such as topology of area (maps), vehicle traffic information. Hence the external information from the other networks viz. maps junctions is the points where special strategies are implemented to make routing decisions. One of the earlier routing protocols that apply the overlay approach is the Greedy Perimeter Coordinator Routing (GPCR) [16]. Furthermore, Greedy Perimeter Stateless Routing at Junction (Gpsrj+) [17], Geographic Source Routing with Directional Forwarding Approaches (DGSR)[18], Geographic Source Routing(GSR) [19], Enhanced GyTAR (E-GyTAR) [20],Greedy Traffic-Aware Routing (GyTAR) [21], The Dynamic Traffic Aware Routing (ITAR) [22],Directional Greedy Routing (DGR) [23], Predictive Directional Greedy Routing (PDGR) [24], Directional Greedy routing protocol (DGRP) [25], and Reliable Directional Greedy Routing (RDGR) [26]are routing protocols belong to this category.

### 4.2.3. Hybrid Position-Based Routing Protocols:

The hybrid position-based routing protocols are a combination of non- DTN and DTN. This category takes the advantage of both DTN and non-DTN categories to solve the disconnect issues of other VANET routing protocols. One of the earlier routing protocols that apply the Hybrid position-based Routing protocols approach is the Hybrid Geographic and DTN Routing with Navigation Assistance in Urban Vehicular Networks environments (GeoDTN+Nav) [27].

## 5. Representative Non-DTN Overlay Position-based Routing Protocols

This section characterizes the unicast non-DTN overlay position-based routing protocols for urban and highway environment, and provides a comprehensive comparison of the surveyed protocols. Figure2 presents the surveyed protocols for urban and highway environment.

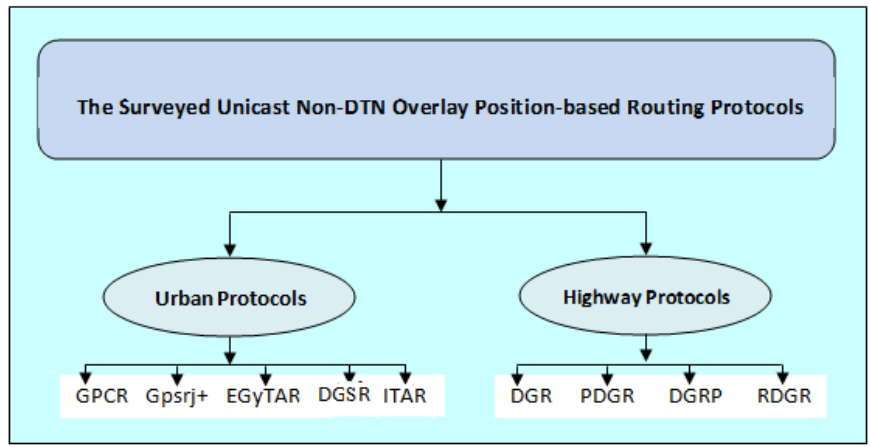

Figure 2. The Surveyed Unicast Non-DTN Overlay Position-based Routing Protocols

### 5.1. Urban Environment

#### 5.1.1. Greedy perimeter coordinator routing (GPCR)

GPCR [16] is a unicast position based routing protocol. GPCR is a vehicle-to-vehicle protocol that is used in urban area. It is a street aware protocol. GPCR algorithm performs two strategies to find the optimal routes. The strategies are: the restricted greedy forwarding on the basis of prior selected route, and the perimeter mode. In GPCR, every node continuously sends beacon packets with their own position and their node id. Also, the beacon includes information about whether the sender is located on a junction or on a street. Each node is aware of position information of its neighbour s, and a node can be considered a coordinator node when it has two neighbour s that are within communication range of each other, but do not list each other as a neighbour . GPCR makes routing decisions on the basis of streets and junctions instead of individual nodes and their connectivity. Also, the coordinator node is responsible for making routing decision without considering digital map. When a node wants to send a packet a location service is used to find out the position of the final destination. Next, the restricted forwarding mode imposes the packet holder node to forward the packets to a node that is located at the junction. Therefore packets should always be forwarded to a Coordinator node on a junction rather than being forwarded across the junction. To determine a coordinator, GPCR algorithm uses two approaches the first one depends on Neighbour Table that were constructed in each node through exploiting beacon messages. The second approach is the Correlation coefficient, neighbour to find the correlation coefficient a node uses its position information and the position information of its immediate nodes. In GPCR, when the restricted greedy fails to forward the data packet i.e. packet fall in void, it uses recovery mode to overcome such problem. In the recovery mode a junction node will be assigned based on the right hand rule to which junction the packet should be forwarded. However,

GPCR suffers from several issues; firstly; because GPCR forwards data packets based on the node density of adjacent roads and the connectivity to the destination, if the density of nodes is low or there is no connectivity to the destination, then the delay time increase. Secondly, the restricted greedy forwarding needs more hops as compared to simple greedy forwarding which also increase delay. Thirdly, the recovery mode also causes extra delay in relaying the packet toward the destination which degrades the throughput of the network. All of those drawbacks degrade the performance of the network.

### 5.1.2. Greedy Perimeter Stateless Routing Junction+ (GpsrJ+)

GpsrJ+ [17] is a unicast position based non-DTN overlay routing protocol. GpsrJ+ is a vehicle-to-vehicle routing protocol which is proposed for the urban environment. It is designed to improve GPCR performance based on minimal modification. GpsrJ+ algorithm consists of two modes; it uses a special form of greedy forwarding i.e packets are greedily forwarded along road segments as close to the destination as possible. When packets reach a void area, the node switches to GpsrJ+ recovery mode. In the recovery mode, packets are greedily backtracked along the perimeter of roads. GpsrJ+ reduces the dependency on junction node also it recovers packets from the void problem by using digital maps. Each node sends a beacon message about its coordinates and the road segments on which its neighbour s are located. GpsrJ+ utilizes two-hop beaconing to predict the next road segment in which the packet should be forwarded toward a destination. A source node and based on the road segments it obtains from its neighbour s' beacon messages, especially its junction neighbour s, the source node pre-computes the segment where it's next hop will be. If the pre-computed advancing segment is similar to the road segment that the source node's furthest neighbour is on, the source node will advance to its furthest neighbour; or else, it will advance to its junction neighbour. However, if the coordinator node has opposite direction from the current forwarding node, it chooses the coordinator node as a next relay one. Accordingly, the major directional decisions are made at junctions. A major advantage of GpsrJ+ protocol is it does not require a costly planarization plan. However, GpsrJ+ algorithm when apply in realistic roads follow a highly complex trajectory even though it uses a simple line trajectory. Further, GpsrJ+ protocol does not have a global view of the network paths.

### 5.1.3. Geographic Source Routing with Directional Forwarding Approaches (DGSR)

DGSR [18] is a unicast position based non-DTN overlay routing protocol. DGSR is a street awareness protocol proposed for the urban environment. DGSR is vehicle-to-vehicle routing protocol. The DGSR protocol is an improvement version of geographic source routing (GSR) protocol [19]. In DGSR, the source nodes which have packet to be forwarded get the position of final destination node via location service. The DGSR protocol algorithm computes the shortest path to the destination using the Dijkstra algorithm. The path is composed of a set of junctions and each packet from source node follows the sequence of junctions to reach final destination. The new improvement to DGSR algorithm is that the forwarding node uses directional greedy forwarding to forward data packets towards the destination. In case forwarded packet stuck at void, DGSR uses carry and forward strategy for local maximum problem. However, DGSR protocol does not take into account the status of link while forwarding. Therefore, in case of high mobility of vehicular nodes, it suffers from packet loss due to link breakage.

### 5.1.4. Enhanced Greedy Traffic Aware Routing Protocol (E-GyTAR)

E-GyTAR [20] is a unicast position based routing protocol. It is a vehicle-to-vehicle traffic aware protocol proposed for the urban environment. E-GyTAR is an enhanced version of GyTAR [21].

It uses location services to get the position of the destination node. Each vehicle maintains a neighbour table, in this table the position, velocity, and direction of each neighbour vehicle are saved. Further, this vehicle's table is updated as a node receives a new beacon message. If the vehicle wants to send a data packet towards the destination then it refers to its neighbour table to find the optimal path. E-GyTAR has also two phases; dynamic junction selection mechanisms based on directional density, and improve greedy forwarding for routing in between junction. E-GyTAR algorithm selects optimal path on the basic of higher number of vehicles moving in the direction of destination. The junction that is nearest to the destination and has the highest score will be selected as a next relay intersection. It uses improved greedy packet forwarding strategy to forward the packet between the junctions. In E-GyTAR, if the packet stuck in the void, then carry and forward technique is used for recovery. The main issue associated with E-GyTAR routing protocol is that it selects junctions based on directional density and ignores non-directional density flows on a multi-lane road. If there is no directional density then data packet will stuck in void area. Thus, E-GyTAR coverts to carry and forward technique which results in more delay that degrade the performance of the network.

### 5.1.4. The Dynamic Traffic Aware Routing (ITAR)

ITAR [22] is a unicast position based routing protocol. ITAR is a traffic aware protocol proposed for the urban environment. ITAR is vehicle-to-vehicle routing protocol. ITAR is an improved version of E-GyTAR. ITAR algorithm improves junction selecting strategy, and recovery strategy used by E-GyTAR. With ITAR, each vehicle has GPS, location service which is Grid Location Service (GLS), digital maps, and Vehicular traffic estimation technique like, IFTIS on board. ITAR selects the junctions that construct the routing path dynamically. When a source generates a packet and each time a packet is at a junction, the neighbour ing junction was scored. The junction with the highest score is selected as the next junction. This selecting mechanism takes into account both the junction's position and the consideration of vehicular traffic on the street. Packets are forwarded through applying an improved greedy forwarding manner as E-GyTAR does. When a packet encounters a local optimum, ITAR algorithm applies a conditional carry and forward approach. The local optimal node should score itself and its neighbour vehicles, and decide whether to carry the packet or forward it based on who is the closest to the destination.

## 5.2 Highway Environment

### 5.2.1. Directional Greedy Routing (DGR)

In [23], the authors propose a protocol named DGR is a unicast position based non-DTN overlay routing protocol. DGR is designed for the highway environments, therefore, unlike city environment, it does not require junction and anchor point selection. DGR is a vehicle-to-vehicle routing protocol. To get the location information of the destination, it requires static maps and location services. Moreover, DGR assumes that the vehicles are aware of their velocity and directional information. DGR uses directional greedy forwarding for sending the packets towards the destination. In case, a source node wants to send a packet to a destination node, DGR algorithm selects a node which is closest to the destination and is moving in the same direction. Furthermore, if the packet holder does not have the vehicle in the direction of the destination then it apply carry and forward recovery mode for certain specified time duration as threshold and try to find out the next vehicle moving in the direction of destination. The Directional Greedy Routing scheme intends to reduce routing loops in the forwarding process. However, may cause more hops and delay.

### 5.2.2. Predictive Directional Greedy Routing (PDGR)

PDGR [24] is a unicast position based non-DTN overlay routing protocol. It is a vehicle-to-vehicle protocol which is proposed for the highway environment. PDGR algorithm able to forward packet to the most suitable next hop based on both current and predicable future situations. Also it has static digital maps and GPS (or DGPS) installed to get its accurate geographical location. PDGR applies Direction First Forwarding (DFF) strategy and Position First Forwarding (PFF) strategy. Hence, the weighted score is computed from these two strategies. The weighted score is estimated for the source node and its current neighbour s, and potential neighbour s. When a source node wants to send packets to a destination, the destination location is known in advance. Each vehicle has the knowledge of its own velocity and direction. Nodes send beacon message continuously with fixed beaconing interval. Thus, each node in the network constructs its own routing table. In the routing table; position information, motion direction and ID of all one-hop neighbour s are maintained. PDGR takes both position and direction into consideration when choosing next hop. The next hop is selected by calculating weighted score of the two metrics. Additionally, a prediction approach is applied by considering the packet carrier's possible future neighbour s make routing more efficient in PDGR. Furthermore, if the packet reaches void i.e. packet holder does not have a vehicle in the direction of the destination then it apply carry and forward recovery mode until it find out the next vehicle moving in the direction of destination. The main drawback of PDGR is the network disconnection problem if the predicted node is selected as next relay-node that it may become out of sender range.

### 5.2.3. Directional Greedy routing protocol (DGRP)

DGRP [25] is a unicast position based routing protocol. It is vehicle-to-vehicle which proposed for urban environment. DGRP uses the two forwarding strategies greedy and perimeter. Further, DGRP algorithm depends on DGR protocol with more improvements. DGRP predicts the position of vehicle neighbouring nodes during the beacon interval. To predict the new positions, DGRP algorithm uses the neighbour s' speed and direction information provided in beacon packets. A source node which has data packet to be forwarded predicts the position of its neighbour's nodes within the beacon interval. Next, it selects the most appropriate next forwarding node. The selection process is based on the neighbour s which is closest to destination or to the next intermediate node. The main drawbacks of DGRP: Each node has to compute the speed and direction of motion, each forwarding node has to predict position of all its neighbour s, and accuracy of position prediction method depends on frequency of change in speed and direction. Such drawbacks result in more delay that affects the overall performance of DGRP.

### 5.2.4. Reliable Directional Greedy Routing (RDGR)

RDGR [26] is a reliable unicast position-based routing protocol. It is vehicle-to-vehicle which proposed for urban environment. Further, RDGR algorithm depends on DGR protocol with more improvements to increase its reliability. The RDGR approach obtains position and movement information of its neighbouring nodes from GPS. The RDGR algorithm uses information of vehicles' position, speed, direction of motion i.e. movement information to predict link stability of their neighbour s. The RDGR algorithm calculates link stability between neighbour nodes in distributed fashion for reliable forwarding of data packet. The packet sender uses neighbour's link stability to select the most appropriate next forwarding node. It uses combination metrics of distance, velocity, direction and link stability to decide about to which neighbour the given packet should be forwarded. The RDGR algorithm incorporates potential score based strategy, which reduces link breaks, enhances reliability of the route and improves packet delivery ratio.

## 6. Performance Evaluations

The applicability of selected routing protocols in urban and highway environments is different. These differences appeared due to the specification of each protocol, such differences have a significant impact on the performance of each routing protocols. This work discusses the properties for the routing protocols and adaptability in each environment. Table 1 and Table 2 present the specification of simulation parameters for both environments. In this section, the work evaluates the performance of the selected protocols in NS-2.34. The researcher compared the performance of DGR, PDGR, DGRP, and RDGR in highway environment and GPCR, Gpsr+, DGSR, EGyTAR, and ITAR in urban environment.

### 6.1. Simulation Parameters

Table 1 shows Simulation Parameters for highway environment, and Table 2 shows Simulation Parameters for Urban environment.

Table 1. Simulation parameters for highway environment

| Simulation | NS2.34 |
|---|---|
| Scenario Area | 5000 x 5000 m |
| Simulation time | 300 Seconds |
| Vehicle Nodes | 200 |
| Transmission Range | 250m |
| Movement model | Random Waypoint |
| Minimum speed value | 70 km/h |
| Maximum speed value | 130 km/h |
| Hello packet size | 12 bytes |
| Hello packet interval | 1.5 second |
| Density between nodes | 5 vehicles every 130 m |
| MAC layer protocol | IEEE 802.11 DCF |
| Traffic Type | CBR /UDP |
| Packet size | 512 bytes |
| Channel bandwidth | 2 Mbps |
| Radio propagation model | Two Ray Ground Model |

Table 2. Simulation parameters for urban environment

| Simulation | NS2.34 |
|---|---|
| Scenario Area | 5000 x 5000 m |
| Simulation time | 300 Seconds |
| Number of nodes | 50-300 |
| Transmission Range | 250m |
| Movement model | Modified Random Waypoint |
| Minimum speed value | 10 km/h |
| Maximum speed value | 50 km/h |
| Hello packet size and interval | 12 bytes and 1.5 second |
| Density between nodes | 1 vehicles every 10 m |
| MAC layer protocol | IEEE 802.11 DCF |
| Traffic Type | CBR/UDP |
| Packet size | 512 bytes |
| Channel bandwidth | 2 Mbps |
| Radio propagation model | Two Ray Ground Model |

### 6.2. Simulation Setup

For urban and highway environments the experiments are carried out in NS-2.34 simulator. The evaluation metrics that are used are Packet Delivery Ratio, Void Problem Occurrence, and Average Hop Count. Those metrics are chosen based on the required solution. It was required to be sure that the data is exchanged between the vehicles in different Hops with the highest delivery, and least void occurrence of the surveyed protocols. The performance of the selected metrics has been evaluated in urban and highway environments. The simulation parameters are as shown in table 1 and table 2. Two different scenarios have been conducted for each environment. For the first scenario the number of vehicles varies from 50 vehicles up to 300 vehicles with fixed speed 40 km\h for urban environment and 90 km\h in highway area. For the second scenario the vehicles' speed varies from 10km\h up to 50 km\h in urban area and from 70km\h up to 130km\h in highway area with constant number of vehicles i.e. 150 nodes. Further, in each scenario 10 pairs of source-destination are randomly selected.

## 7. Experimental Results and Discussion

7.1 **Highway Environment**

**7.1.1. Various Numbers of Nodes**

In the first scenario for highway area, a variance number of vehicles; 50, 100, 150, 200, 200, 250 and 300 are used and all vehicles move with 90 km/h. Figure 3 shows the impact of node density on the surveyed routing protocols based on the selected performance metrics.

Figure 3(a) shows the packet delivery ratio for DGR, PDGR, DGRP and RDGR; researcher can notice that the packet delivery ratio increases as number of node increases for all protocols. This is because more number of nodes provides opportunity to select suitable neighbour node. The packet delivery ratios of DGR and PDGR protocols are slightly the same. And PDGR slightly outperform DGR in packet delivery ratio because of its prediction algorithm. DGRP outperforms DGR and PDGR because of its new features. Further, RDGR outperforms all other protocol because RDGR algorithm calculates the potential score for packet forwarding node and its neighbour nodes to make forwarding decision.

Figure 3(b) shows the average number of hops for DGR, PDGR, DGRP and RDGR; researcher can notice that the average number of hops increases as number of node increases for all protocols. This is because more number of nodes provides opportunity to reach destination with more hops. DGR and PDGR protocols haves lightly the same hops count, DGRP has less hops count than DGR and PDGR, and RDGR has the least average number of hops compared to the other protocols because of its selection criterion of the next forwarding node.

Figure 3(c) shows the void problem occurrence rate for DGR, PDGR, DGRP and RDGR; researcher can notice that the void problem occurrence rate decreases as number of node increases for all protocols. This is because more number of nodes provides opportunity to make correct routing decision. RDGR has the least void problem occurrence rate compared to the other protocols because of its selection criterion of the next forwarding node.

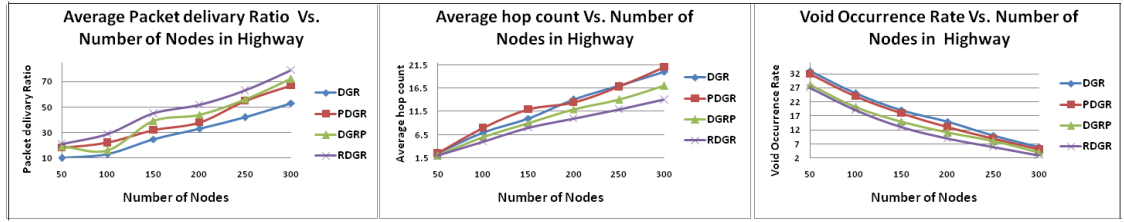

Figure 3. The impact of node density on the surveyed routing protocols based on the selected performance metrics

### 7.1.2. Variety Movement Speeds

In the second scenario for highway area, a fixed number of 150vehicles, and a variety movement speeds; 130, 120, 110, 100, 90, 80, and 70 Km/h were deployed. Figure 4 shows the impact of mobility on surveyed routing protocols based on the selected performance metrics.

Figure 4(a) shows the packet delivery ratio for DGR, PDGR, DGRP and RDGR; researcher can notice that the packet delivery ratio decreases as the speed of vehicles increases for all protocols. This is because the increment in speed results more inaccuracy in position information which increase packet loss opportunity, also it results in decreasing in link stability. RDGR outperforms all other protocol because RDGR algorithm can manage the accuracy of position information and link stability problems better than other protocols. Further, the packet delivery ratios of DGR and PDGR protocols are slightly the same. Also, DGRP outperforms both DGR and PDGR Protocols.

Figure 4(b) shows the average number of hops for DGR, PDGR, DGRP and RDGR; the results indicate that the average number of hops increases as the speed of vehicles increases for all protocols. The increment in hops number result more delay to make routing decision at each hop which results in bad performance of the routing protocol. RDGR showed the least void problem occurrence rate compared to the other protocols.

Figure 4(c) shows the void problem occurrence rate for DGR, PDGR, DGRP and RDGR; researcher can notice that the increment in vehicles' speeds results in higher occurrences of void problem. This is because the increment in speed results more inaccuracy in position information which of high opportunity to forward packet to wrong intermediate node, also it results in decreasing in link stability that may results in link breakage between the communicating nodes. RDGR showed the least void problem occurrence rate compared to the other protocols.

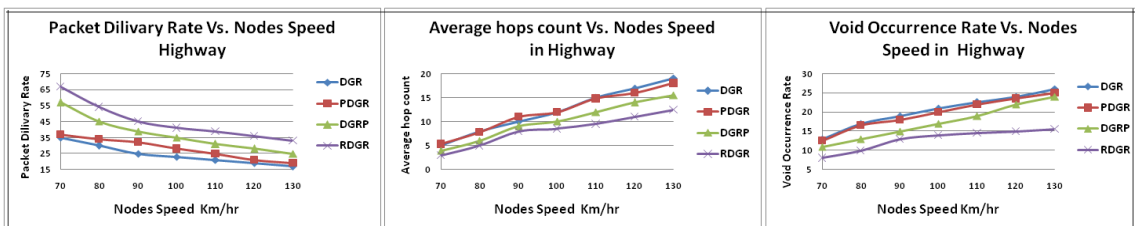

Figure 4. The impact of mobility on surveyed routing protocols based on the selected performance metrics.

### 7.2 Urban Environment

#### 7.2.1 Various Numbers of Nodes

In the first scenario for urban area, a variance number of vehicles; 50, 100, 150, 200, 200, 250 and 300 are used and all vehicles move with 50 km/h. Figure 5 shows the impact of node density on the surveyed routing protocols based on the selected performance metrics.

Figure 5(a) shows the packet delivery ratio for GPCR, Gpsrj+, DGSR, EGyTAR and ITAR; researcher can notice that the packet delivery ratio increases as number of node increases for all protocols. This is because more number of nodes provides opportunity to select suitable neighbour node. Since more nodes are involved in forwarding process, undeliverable packets may create a loop and the same route formed by the same nodes in the same many hops is visited again. However, ITAR protocol achieved highest packet delivery ratio and shows a steady performance comparing to other protocols.

Figure 5(b) shows the average number of hops for GPCR, Gpsrj+, DGSR, EGyTAR and ITAR; researcher can notice that the average number of hops increases as number of node increases for all protocols. This is because more number of nodes provides opportunity to reach destination with more hops. However, ITAR protocol achieved lowest number of hops values and shows a steady performance comparing to other protocols.

Figure 5(c) shows the void problem occurrence rate for GPCR, Gpsrj+, DGSR, EGyTAR and ITAR; researcher can notice that the void problem occurrence rate decreases as number of node increases for all protocols. This is because more number of nodes provides opportunity to make correct routing decision. However, ITAR protocol achieved lowest number of void problem occurrence rate and shows a steady performance comparing to other protocols.

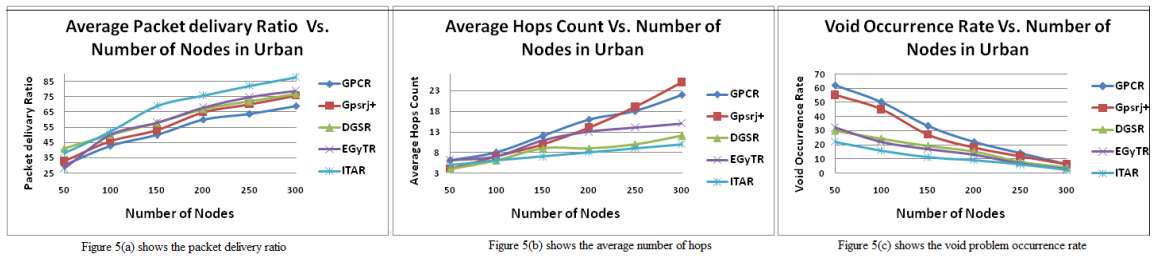

Figure 5(a) shows the packet delivery ratio    Figure 5(b) shows the average number of hops    Figure 5(c) shows the void problem occurrence rate

Figure 5. The impact of node density on the surveyed routing protocols based on the selected performance metrics.

### 7.2.2 Variety Movement Speeds

In the second scenario for urban area, a fixed number of 150vehicles, and a variety movement speeds; 50, 40, 30, 20, and 10 Km/h were deployed. Figure 6 shows the impact of mobility on surveyed routing protocols based on the selected performance metrics.

Figure 6(a) shows the packet delivery ratio for GPCR, Gpsrj+, DGSR, EGyTAR and ITAR; researcher can notice that the packet delivery ratio decreases as the speed of vehicles increases for all protocols. This is because high mobility results in frequent topology changes that affect the performance of the routing protocols. Frequent topology changes decrease in link stability between vehicles neighbour s .ITAR shows better in packet delivery ratio, due to the enhanced recovery mode it adopts. Further, wrong carrying decision of other protocols will contribute to the packet loss rate.

Figure 6(b) shows the average number of hops for GPCR, Gpsrj+, DGSR, EGyTAR and ITAR; we can notice that the average number of hops increases as the speed of vehicles increases for all protocols. The topology changes are proportional to speeds of the nodes that are also proportional to stale position information of vehicle. Choosing a next relay-node with stale information results in using a stale rout that incurs more hops to be traversed by the routed packet. ITAR shows better performance in hops count compared to the other protocols. This trend occurs because ITAR algorithm explicitly considers vehicle mobility characteristics to make forwarding decision.

Figure 6(c) shows the void problem occurrence rate for GPCR, Gpsrj+, DGSR, EGyTAR and ITAR; researcher can notice that the void problem occurrence rate increases as the speed of vehicles increases for all protocols. The results indicate that ITAR performance is better in compared to other protocols. The reason is the modification done in ITAR protocol, in which the data packet can be received by the next relay node. This improvement was achieved at the expense of extra delay.

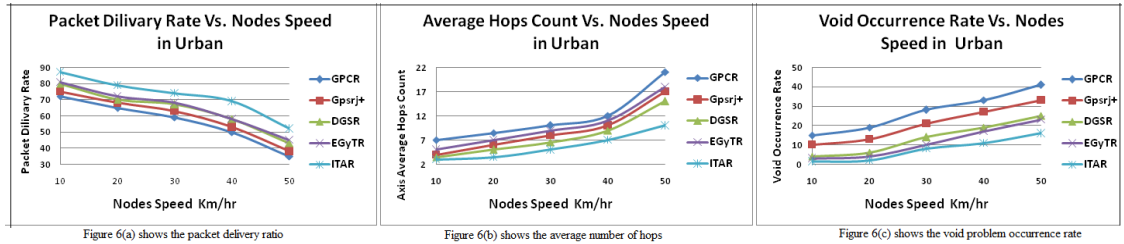

Figure 6. The impact of mobility on surveyed routing protocols based on the selected performance metrics.

## 8. Conclusion and Future Work

The VANETs can be deployed in three different environments: highways, city and rural. In rural areas, network is sparsely populated and hence vehicle to vehicle communication face the problem of instability of links due to frequent topology change. The city environment contains junctions and obstacles in the form of high rise buildings. The highways have relatively better communication as most of the vehicle move along the same path; also, it contains no obstacles. In VANETs, the design of efficient routing protocol for effective vehicular communications poses a series of technical challenges. The need to design efficient and scalable routing protocols for VANET makes position-based routing attractive choice. To accomplish this goal, this article presents a vast number of routing protocol. The surveyed protocols are intended as an aid in the difficult protocol comparison and selection task. The starting points of an analysis are the particular characteristics of the VANET network. Once the environment where it should be set up is established, the focus can move to the most stringent demands of the routing protocol and on it's the desired characteristic. The protocol choices can therefore be narrowed down to just a few potential candidates from the already investigated position-based routing protocols that can satisfy VANET needs. The future work direction will to develop a routing strategy that is applicable to highways and city environments. That protocol will fulfil the requirement of routing in vehicular Ad-hoc networks.